\documentclass[12pt]{article}
\usepackage[utf8]{inputenc}
\usepackage[T1]{fontenc}

\usepackage[nocompress]{cite}
\usepackage{IEEEtrantools,stackengine}
\stackMath

\usepackage{authblk}
\usepackage{fullpage}
\usepackage{amssymb,amsmath}

\usepackage{siunitx}
\usepackage{csquotes}
\usepackage{comment}

\usepackage{xspace}
\newcommand\ie{i.e.\xspace}
\newcommand\eg{e.g.\xspace}

\usepackage{booktabs}
\usepackage{longtable}
\usepackage{multirow}
\usepackage{afterpage}

\usepackage[dvipsnames]{xcolor}

\usepackage[
    colorinlistoftodos,
    textsize=footnotesize,
        ]{todonotes}

\definecolor{darkgreen}{rgb}{0.0, 0.5, 0.0}

\usepackage{setspace}
\usepackage[unicode=true]{hyperref}
\hypersetup{breaklinks=true,
            bookmarks=true,
            colorlinks=true,
            pdfborder={0 0 0},
            allcolors=blue}
\usepackage[%
  sort&compress
]{cleveref}
  \Crefname{appendix}{Supplement}{Supplements}

\usepackage{times}
\usepackage{enumerate}
\usepackage{float}

\usepackage{subfig}
\usepackage{graphicx}

\newcommand{\figletter}[1]{{{\fontfamily{\sfdefault}\selectfont \textbf{#1}}}}

\usepackage{rotating}
\usepackage{tabularx}
\usepackage{dcolumn}
\usepackage{pdflscape}
\usepackage{rotating}

\usepackage{array}

\usepackage{sectsty}
  \subsubsectionfont{\normalfont\itshape}

\makeatletter
\renewcommand{\fps@figure}{H}         
\renewcommand{\fps@table}{H}         
\makeatother

\allowdisplaybreaks

\usepackage{eurosym}

\usepackage{ntheorem}
\theoremseparator{:}

\usepackage{enumitem}

\begin{document}

\title{\centering\LARGE\singlespacing Gender-based discrepancies in the algorithmic delivery of political ads on social media}










\renewcommand\Affilfont{\fontsize{9}{10.8}\selectfont}

\author[1,*]{Dominik Bär}
\author[2,3,*]{Francesco Corso}
\author[3]{Gianmarco De Francisci Morales}
\author[1]{Stefan Feuerriegel}
\author[2]{Francesco Pierri}

\affil[1]{LMU Munich \& Munich Center for Machine Learning (MCML), Munich, Germany}
\affil[2]{Politecnico di Milano, Dipartimento di Elettronica, Informazione e Bioingegneria, Milan, Italy}
\affil[3]{Intesa Sanpaolo Innovation Center, Turin, Italy}
\affil[*]{Equal Contribution}
\date{}

\maketitle

\clearpage
\begin{abstract}\normalfont
\noindent Social media has become a key channel for political advertising during election campaigns. However, algorithmic biases in the delivery of these ads may distort the public’s exposure to political messaging. 
This can hinder citizens' ability to make informed choices and undermine equal access to political discourse, raising concerns about the integrity of electoral processes. 
In this study, we examine gender-based discrimination in the delivery of political ads during the 2024 European Parliament elections. 
Using a large-scale dataset of over \num{110000} ads from 453 political parties and 968 candidates that generated over 7 billion impressions across 25 EU countries, we find that men were significantly more likely to be shown ads from populist and far-right parties than women---even after accounting for ad content, platform-level competition, and targeting strategies. 
All else equal, ads by populist parties reach, on average, a 6 percentage point higher male share. 
Such imbalances restrict the ability of parties to reach diverse audiences and prevent voters from engaging equally with the full range of political viewpoints. 
This pattern is particularly concerning given that far-right and populist ads may reinforce political polarization and widen existing gender gaps in political engagement. 
Our findings underscore the need for platforms and policymakers to audit algorithmic ad delivery in political campaigns on social media and to implement safeguards that ensure fairness and protect democratic processes.
\end{abstract}

\flushbottom
\maketitle
\thispagestyle{empty}

\sloppy
\raggedbottom



\clearpage
\section*{Main}
\label{sec:introduction}

Social media has become an essential tool for political advertising during election campaigns \cite{Bright.2020, Fowler.2021, Fossen.2021, Mallipeddi.2021, Bar.2024b, Bar.2025}. 
Globally, political advertisers published over 2.5 million ads on Facebook and Instagram between August 2020 and December 2022 \cite{Votta.2024}. 
In the United States alone, advertisers spent over USD~619 million on digital ads during the 2024 election cycle \cite{Vandewalker.2024}. 
Similarly, in Europe, political ads on social media accumulated over 1.1 billion impressions during the 2021 German federal election \cite{Bar.2024b} and are considered a key campaign tool by candidates \cite{GLES.2022, Capozzi.2021}. 
Given the influence that political advertising on social media has on electoral outcomes \cite{Hager.2019, Coppock.2022, Aggarwal.2023, Bar.2025}, ensuring fairness and accountability in social media advertising is critical to maintaining the integrity of democratic processes.

Political ads on social media are delivered through automated algorithms that optimize ad placement based on user data. Advertisers define a target audience, ad timing, and campaign budget, while the algorithm determines which users are most likely to engage with the content \cite{Ali.2019}. 
However, these algorithms are typically opaque, proprietary, and controlled by private companies, thus limiting both public oversight and transparency \cite{Ali.2019, Lambrecht.2019, Imana.2021, Ali.2021}. 
This lack of transparency has raised concerns regarding fairness and accountability in electoral contexts \cite{Edelson.2019, Edelson.2020, Ali.2021, LePochat.2022, Bar.2024b, Capozzi.2020}. 
At an individual level, privacy concerns emerge because algorithmic delivery may rely on personal or sensitive user information \cite{Korolova.2011,Auxier.2020, Cabanas.2021}. 
At a societal level, algorithms may exhibit biases that distort the audience of an ad \cite{Lambrecht.2019, Speicher.2018, Ali.2019, Ali.2021, Imana.2021, Ali.2023, Sankaranarayanan.2023, Imana.2024}. 
In political advertising, for example, ads may be disproportionately shown to men, even if the advertising party does not specifically target a male audience. 
This skew may narrow the range of voters that parties are able to reach and reduce citizens’ equal exposure to competing political perspectives. The concern is amplified by the fact that far-right and populist advertising can fuel polarization and further entrench gendered patterns of political engagement \cite{Davis.2025}.

Existing literature highlights that algorithmic ad delivery on social media often leads to audience distortions \cite{Lambrecht.2019, Speicher.2018, Ali.2019, Ali.2021, Imana.2021, Ali.2023, Sankaranarayanan.2023, Imana.2024}. 
These distortions are observed across various types of advertising content, including job ads \cite{Lambrecht.2019, Ali.2019, Imana.2021}, education ads \cite{Imana.2024}, climate communication \cite{Sankaranarayanan.2023}, and political ads \cite{Speicher.2018, Ali.2021, Imana.2021, Ali.2023, Imana.2024}, often reflecting demographic (e.g., age, gender) \cite{Lambrecht.2019, Sankaranarayanan.2023, Ali.2019} as well as sociocultural biases (e.g., race, political orientation) \cite{Ali.2019, Ali.2021}. 
Existing studies on political advertising primarily focus on overall discrepancies, such as whether ads target ideologically aligned or diverse audiences, often examining these issues within the context of a single country \cite{Ali.2021}. 
However, it remains unclear how these distortions generalize to other advertising markets and whether the characteristics of advertisers (e.g., the political orientation of a party) influence algorithmic ad delivery, and lead to audience biases across different political groups \cite{Capozzi.2020}. 
In light of these issues, this study extends our current understanding of algorithmic biases by focusing specifically on political ads by populist and non-populist parties in the 2024 European Elections, offering a cross-national analysis across 25 EU countries.


Populism has grown considerably worldwide, with prominent examples including the election of Donald Trump in the United States and the rise of populist parties across the European Union. Populism is often viewed as a threat to democracy \cite{levitsky2018populism, mounk2019dictators, Ruth.2023}, due to its role in promoting anti-democratic narratives \cite{Capozzi.2023}, spreading misinformation, deepening political polarization, and undermining democratic institutions \cite{Vachudova.2021}. 
Recent evidence suggests that populist parties benefit particularly from political ads on social media \cite{Capozzi.2023, Bar.2024b}. 
For instance, during the 2021 German federal election, the populist AfD (Alternative für Deutschland) reached considerably more men with ads published on Facebook and Instagram compared to non-populist parties \cite{Bar.2025}. 
These dynamics raise important questions about whether algorithmic ad delivery contributes to audience distortions in ads delivered by populist vs non-populist parties on social media.



There are several factors that may explain biases in algorithmic ad delivery.
For instance, populist parties often employ specific rhetoric \cite{Dai.2022}, which is typically divisive, emotionally charged, and negative \cite{ernst2019populists}. 
Such rhetoric may resonate particularly well with certain user groups, leading to distortions in ``whom'' algorithmic delivery shows populist vs. non-populist ads to. 
Political parties are also commonly associated with distinct demographic and socio-cultural groups \cite{Winter.2010, Zingher.2018, Guth.2021}. 
In the United States, for example, the Republican party is often associated with masculinity, while the Democratic party is more likely to be linked to femininity \cite{Winter.2010}. Similarly, populist parties in Europe tend to attract a predominantly male, less educated electorate \cite{Guth.2021}. 
Algorithmic ad delivery may learn and amplify these associations, even if parties do not explicitly target specific audiences. 
These audience skews may weaken parties’ capacity to communicate across demographic groups and limit voters’ opportunities to encounter the full breadth of political messages. This is particularly important because far-right and populist campaigns may sharpen polarization and aggravate existing gender inequalities in political participation.


In this study, we examine gender-based discrepancies in the algorithmic delivery of political advertisements on social media during the 2024 European Elections, comparing populist and non-populist parties. 
Specifically, we use regression analysis to compare the gender distribution of audiences for political ads from populist and non-populist parties published on Facebook and Instagram. 
Importantly, we focus on ads that do not explicitly target a specific gender, and we account for other factors, such as ad content, platform-level features, and targeting strategies beyond gender. 
By doing so, we isolate the relationship between algorithmic ad delivery and gender-based discrepancies in the audience of populist vs. non-populist ads on social media.

Our analysis is based on a large-scale dataset consisting of over \num{110000} ads from 453 political parties and 968 candidates across 25 EU member states, published on Facebook and Instagram. 
These ads generated over 7 billion impressions and incurred a total of over EUR 23 million in ad spending. 
Overall, we provide a cross-national perspective on discrepancies in the algorithmic delivery of political ads during one of the largest democratic elections globally, involving over 373 million eligible voters \cite{Reuters.2024}.

We find that ads from populist parties were significantly more likely to be shown to men than women---accounting for ad content, platform-level competition, and targeting strategies beyond gender. 
All else equal, ads by populist parties reach, on average, a 6 percentage point higher male share.
Such uneven delivery can reduce the demographic breadth of political communication and create unequal exposure to electoral viewpoints. This pattern raises particular concern because far-right and populist messages may strengthen polarization and widen already existing gender gaps in political engagement.
Our findings underscore the urgent need for platforms and policymakers to audit the algorithmic delivery of political ads on social media and implement safeguards to ensure fairness and protect democratic processes.

\clearpage
\section*{Results}
\label{sec:results}

\subsection*{Advertising landscape during the 2024 European Elections}

The 2024 elections to the European Parliament constitute one of the largest democratic processes globally. 
A total of 720 Members of the European Parliament (MEPs) were elected to represent over 450 million citizens across the 27 member states of the European Union (EU). 
The European Parliament serves as the legislative branch of the EU, playing a central role in shaping agricultural policies, economic integration, and trade agreements. 
The elections were held between June 6 and June 9, 2024, reflecting the diversity of national electoral traditions across member states. 
This electoral context offers the opportunity to examine political advertising at scale within a single supranational election, enabling a cross-national analysis over all EU member states. 
Additional details on the political system of the EU and the 2024 European elections are in \Cref{supp:politics_EU}.

Our analysis relies on a large-scale sample of over \num{110000} political ads published by 968 candidates and 453 parties across the 25 member states of the European Union on Facebook and Instagram during the 2024 European Election. 

We have collected these ads via the Meta Ad Library. 
Since France and Portugal prohibit political advertising in the 12 and 2 weeks leading up to an election, respectively, we have removed those countries from our analysis. 
Our observation period includes the three months leading up to the election (\ie, March 9 to election day on June 9, 2024), which resembles the main campaigning period. In Germany, for instance, candidacies were officially announced on March 18.
To identify the political actor behind each ad, we matched ads to candidates and parties using the advertiser’s page name and sponsor information, combining automated string-matching procedures with manual verification (see Methods).
Overall, advertisers have spent more than 23 million EUR on the ads in our sample, generating a total of over 7 billion impressions and 5 billion unique views. 
An impression is counted each time an ad is rendered on a user's screen, whereas a unique view is counted for each unique user who has seen the ad. 
In the following, we focus on unique views to measure ad exposure, as Meta provides exact values for unique views, but only ranges for impression counts. 

\subsection*{Gender-based discrepancies in the algorithmic delivery
of political ads on social media}
\subsubsection*{Descriptive analysis}

We aim to evaluate gender-based discrepancies in the delivery of political ads by populist and non-populist parties on social media. 
To do so, we rely on the PopuList~\cite{Rooduijn.2024} to classify whether ads were published by populist or non-populist parties.
Importantly, we classify ads by candidates according to their party affiliation, and refer to all ads associated with a populist candidate or party as ads published by the populist party. 
Our sample includes \num{38780} ads from populist parties, which account for 25\% of the total ad spending in our data. 
Populist ads have garnered more than 1.5 billion impressions (~21\% of all impressions) and 1.1 billion unique views (20\% of all unique views).

We first examine how unique views for political ads are distributed across male and female users. 
Overall, 54.1\% of all views are by male users, while only 45.9\% are by female users. 
This result suggests a tendency for political ads to be more likely to be shown to male users. 

Next, we analyze the excess share of unique views by male users (EMS) for ads from populist and non-populist parties across the European Union. 
We define the excess share as the difference between the share of unique views by male users and the share of unique views by female users. 
A value near zero indicates a gender-balanced audience, while positive values reflect a higher proportion of male viewers, and negative values indicate a female-skewed audience.

\begin{figure}[H]
\vspace{-1cm}
{\raggedright\figletter{a)}}

\begin{minipage}[b]{\textwidth}

    \centering
    \hspace{1.5cm}
    \includegraphics[width=.62\textwidth]{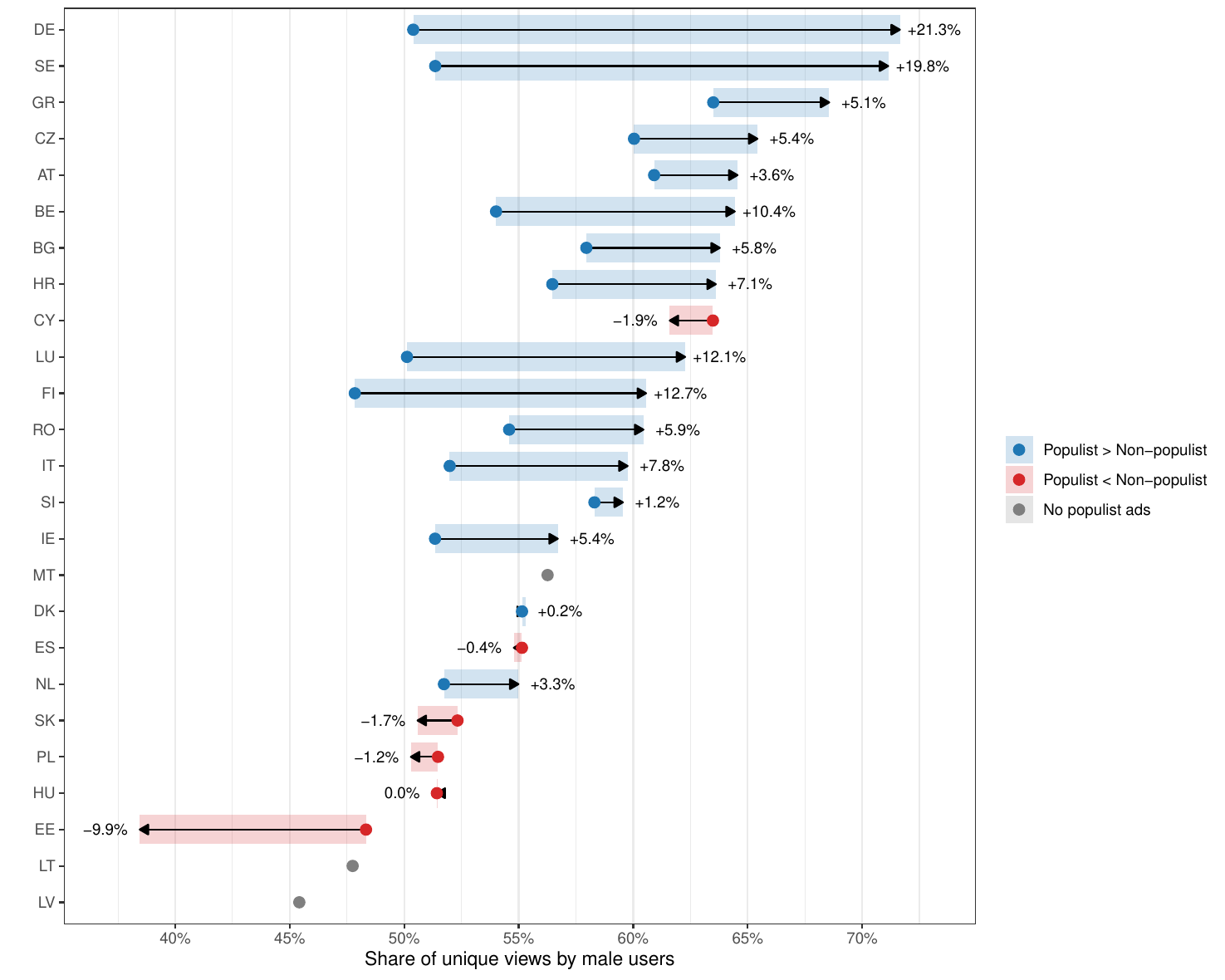}
\end{minipage}
{\raggedright\figletter{b)}}

\begin{minipage}[b]{\textwidth}
    \centering
    \includegraphics[width=\textwidth]{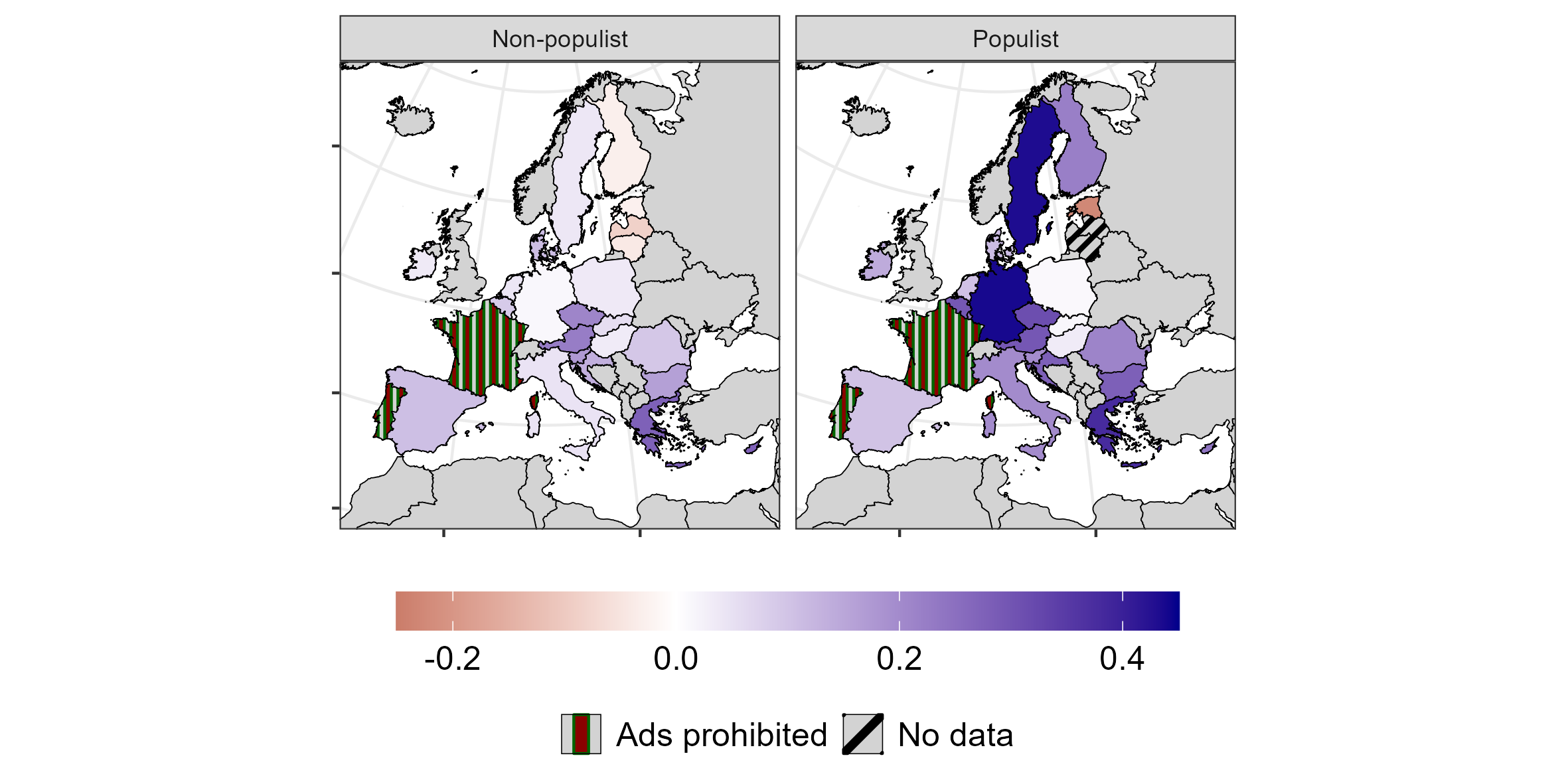}
\end{minipage}

\caption{\footnotesize \textbf{Gender-based discrepancies for populist vs. non-populist social media ads during the 2024 European elections.} \figletter{a)}~The figure illustrates gender-based discrepancies in the audience of political ads across EU member states. Dots represent the share of unique views by male users for ads by non-populist parties. Arrows indicate the excess share of unique views by male users for populist relative to non-populist parties. Countries in which populist parties attract a higher share of male viewers than non-populist parties are shaded in blue; those where the reverse is true are shaded in red. Countries without any populist ads are shown in grey. \figletter{b)}~The map displays the excess share of unique views by male users for political ads published on Facebook and Instagram by populist and non-populist parties across EU member states. A value near zero indicates a gender-balanced audience, while positive values reflect a higher proportion of male viewers. Darker blue shades represent countries where ads from populist parties received significantly more unique views by male users compared to those from non-populist parties. Party classifications are based on the PopuList database~\cite{Rooduijn.2024}. France and Portugal (hatched in red) prohibit political advertising during election campaigns. Latvia and Lithuania (hatched in black) had no ads by populist parties.}
\label{fig:map_excess_male_populist}

\end{figure}

\Cref{fig:map_excess_male_populist}\textbf{a} presents the excess share of unique views by male and female users for populist vs. non-populist ads across all countries in the European Union. 
We observe that populist parties tend to receive more views from male users.
Specifically, ads by populist parties surpass the share of unique views by male users compared to non-populist ads in 16 out of 22 member states that featured populist ads (min = +0.2\%, max = +21.3\%). 
In general, ads from non-populist parties are more likely to reach a gender-balanced audience, with an average excess male share of +7.3\% (see dots in the figure). 

We also examine regional patterns in the excess share of unique views by male users across EU member states (\Cref{fig:map_excess_male_populist}\textbf{b}). 
The highest excess shares of male viewers are observed in Germany (+21.3\,\%), Sweden (+19.8\,\%), and Finland (+12.7\,\%). 
In contrast, populist ads attract more unique views from female than male users in Estonia ($-$9.9\,\%), Cyprus ($-$1.9\,\%), and Slovakia ($-$1.7\,\%). 
Ads by non-populist parties in the Baltic states (i.e., Estonia, Latvia, and Lithuania) tend to receive comparatively fewer views from male users than in other EU countries. 
In contrast, Scandinavian countries (i.e., Sweden and Finland) and several central European states (including Belgium, Germany, the Czech Republic, and Austria) display a larger gender-based discrepancy among viewers of populist ads, with a stronger skew toward male audiences.

\newpage

\subsubsection*{Regression analysis}

To evaluate gender-based discrepancies in the delivery of political ads by populist and non-populist parties on social media, we focus on political ads that do not explicitly target users based on gender. 
The rationale behind this approach is that such ads should be equally likely to be shown to both men and women, once we account for four key factors that may influence the audience of political ads on social media: (1)~ad content, (2)~ad characteristics, (3)~targeting strategies, and (4)~country level heterogeneities.

Our observational analysis shows that ads published by populist parties are more likely to reach a male-dominated audience. 
To evaluate whether this gender-based discrepancy is driven by algorithmic bias, we use regression analysis to isolate the key factors influencing the audience of political ads on social media. 

We use a linear regression model to study the relationship between the factors from above and gender-based discrepancies in ad delivery by populist versus non-populist parties. In this model, the excess share of unique views by male users is regressed on whether the ad was published by a populist party, controlling for a comprehensive set of additional variables.

Our controls account for the first three key factors influencing the delivery of political ads on social media. (1)~We control for \emph{ad content} by including both textual and visual features in our regression model. (2)~We include controls for \emph{ad-level characteristics}, such as platform-level competition, timing (weekday and ad duration), the platform on which the ad was published (Facebook, Instagram, or both), and whether the ad was published by a candidate or a party. (3)~We account for the ad’s demographic targeting based on age groups (\ie, 18--24, 25--34, 35--44, 45--54, 55--64, and 65+). In addition, we include country fixed effects to account for unobserved national-level heterogeneity, such as differences in regulation regarding political advertising on social media. By controlling for these factors, our model aims to isolate the association between gender-based discrepancies and the role of algorithmic bias in the delivery of political ads on social media platforms. Details are in the \nameref{sec:methods}.

\begin{figure}[H]
    \centering
    \includegraphics[width=0.8\textwidth]{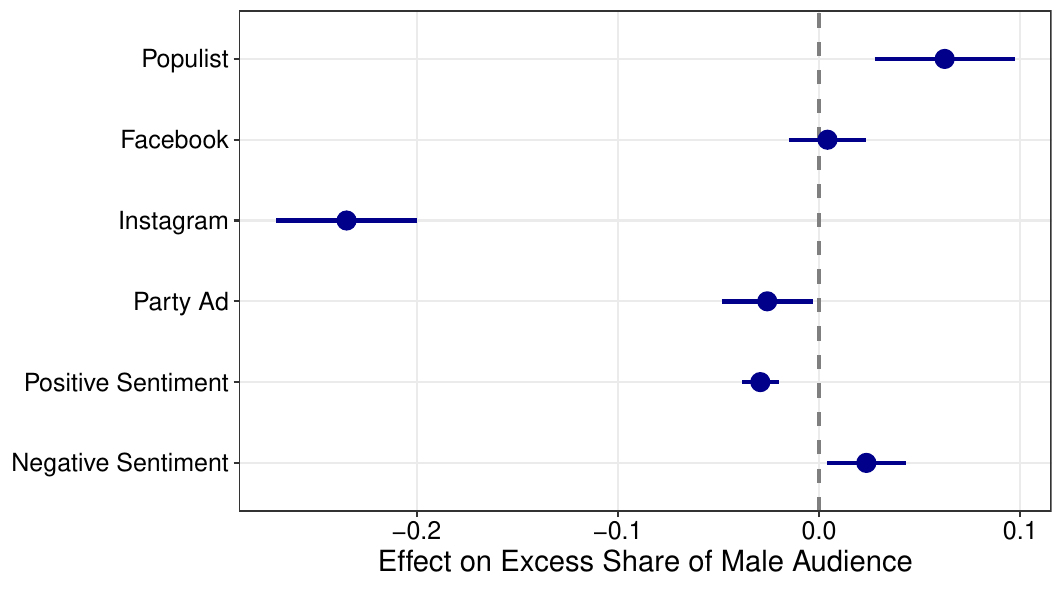}
    \caption{\textbf{Estimation results for our linear regression model, where our dependent variable is the excess share of unique views by male users.} Standard errors are clustered at the national party level to account for potential dependencies among ads published by the same party. 
    Adjusted $R^2$ = 0.245, Number of observations = \num{110468}. Whiskers indicate $\alpha = 95\%$ confidence intervals.}
    \label{fig:coefficient_plot_table}
\end{figure}
\begin{figure}[H]
    \centering
    \includegraphics[width=0.4\textwidth]{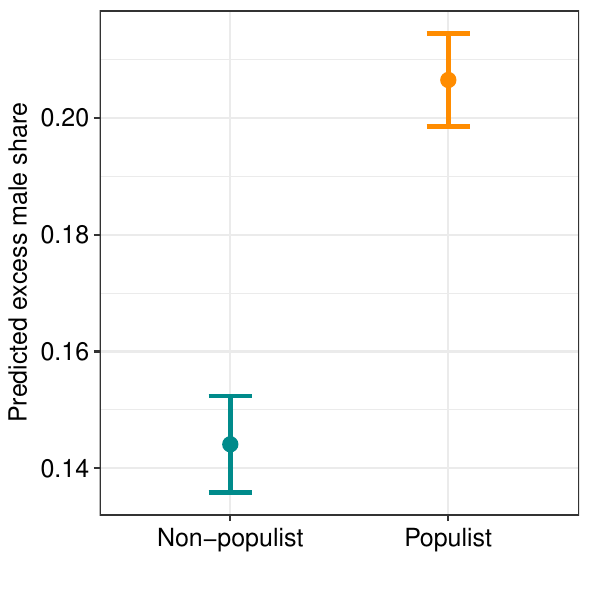}
    \caption{\textbf{Prediction of whether an ad was published by a populist vs. non-populist party on excess share of unique views by male users.} The figure shows the predicted marginal effect of whether an ad was published by a populist vs. non-populist party on the excess share of unique views by male users. Effect sizes are computed by averaging the effects over the observed values of the variables in our model. Whiskers indicate $\alpha = 95\%$ confidence intervals.}
    \label{fig:results_regression}
\end{figure}

The estimated coefficients of the regression model are reported in \Cref{fig:coefficient_plot_table}. 
We find a positive and statistically significant association between the excess share of unique views by male users and an ad being published by a populist party (coefficient: $0.062$, $\text{SE} = 0.0020$, $t = 30.49$, $P<0.001$, 95\,\%~CI~$[0.058, 0.066]$). This result suggests that, all else equal, ads published by populist parties are associated with a 6.2 percentage point higher excess share of unique views by male users, on average. \Cref{fig:results_regression} shows the predicted excess male share of political ads. Ads from populist parties are thus significantly more likely to reach male-dominated audiences. Given that we control for (1)~ad content, (2)~ad-level characteristics, and (3)~targeting strategies, this association may reflect biases in the algorithmic delivery of political ads on Facebook and Instagram.

\subsubsection*{Robustness checks}

To ensure the robustness of our results, we further perform an extensive series of checks regarding (1)~alternative definitions of our dependent variable (\eg, we use the share of unique views by male users as an alternative dependent variable), (2)~model specification (\eg, we account for the baseline gender distribution in a country), (3)~data sample (\eg, we check different campaign periods), (4)~modeling approach (\eg, we use propensity score matching to compare ads by populist vs. non-populist parties and random effects for countries instead of fixed effects) and (5)~ablating media and targeting covariates. Across all checks, our findings remain robust. That is, we find a statistically significant gender-based discrepancy favoring male audiences for populist ads published on Facebook and Instagram. Details are in \Cref{supp:robustness_checks}.

\subsection*{The role of political extremism}

Populist parties encompass a broad range of political ideologies that can vary considerably, especially across national contexts \cite{Capozzi.2023}. Populist movements have emerged from both ends of the political spectrum \cite{Vachudova.2021}. 
For example, left-leaning populist parties have gained prominence in Southern Europe, whereas populist parties in Central and Eastern Europe are more commonly aligned with the political right \cite{Vachudova.2024}. Despite these ideological differences, populist parties often exhibit commonalities \cite{Capozzi.2023} and tend to operate at the fringes of the political spectrum. These patterns suggest that algorithmic ad delivery may not only differ between populist and non-populist parties, but also between the ideological extremes of the spectrum, \ie, between far-right and far-left parties. To identify the ideological orientation of each party in our data, we use the classification provided by the PopuList database~\cite{Rooduijn.2024}. 


\subsubsection*{Descriptive analysis}

\begin{figure}[H]
\vspace{-0.5cm}
{\raggedright\figletter{a)}}

\begin{minipage}[b]{\textwidth}
    \centering
    \includegraphics[width=.6\textwidth]{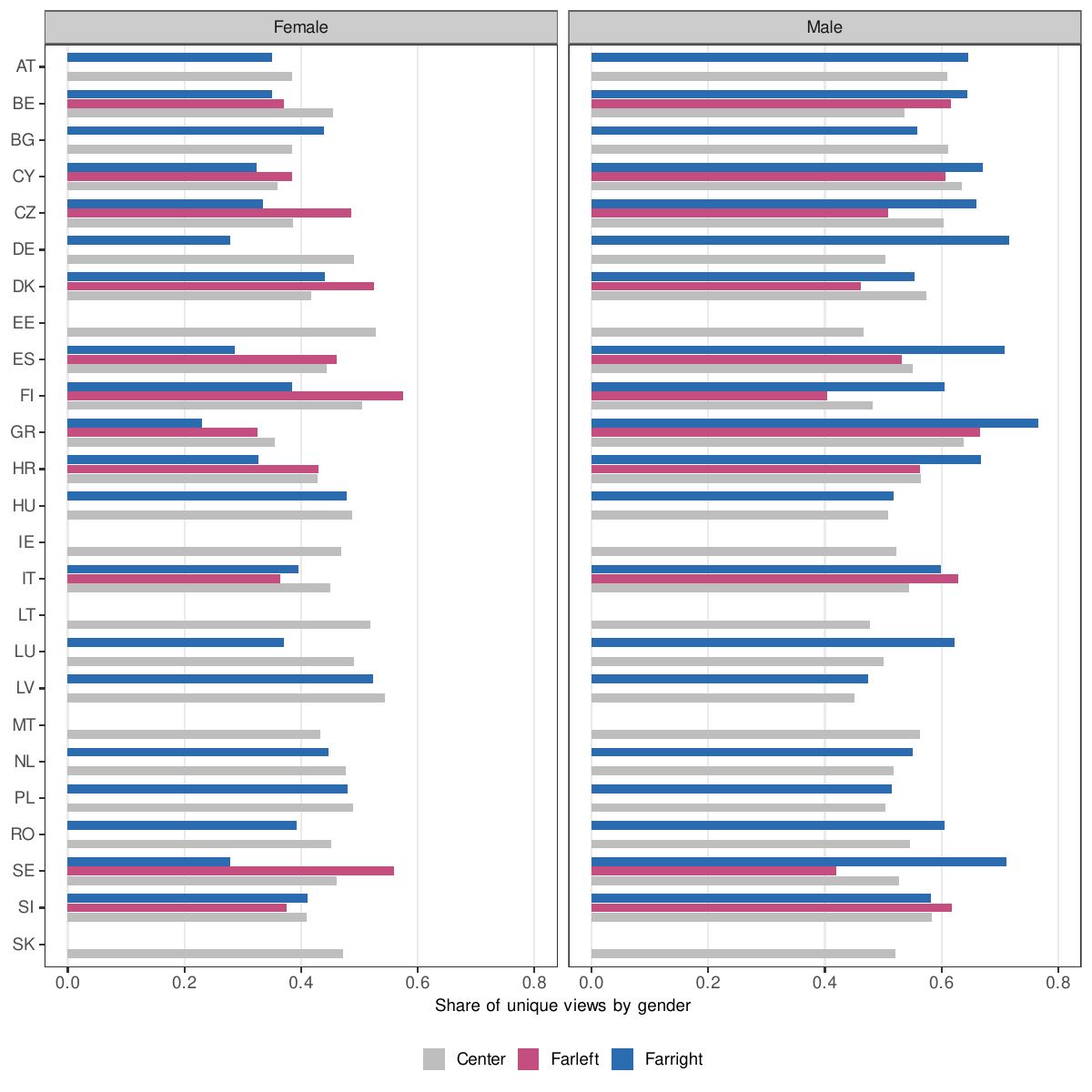}
\end{minipage}
{\raggedright\figletter{b)}}

\begin{minipage}[b]{\textwidth}
    \centering
    \includegraphics[width=.7\textwidth]{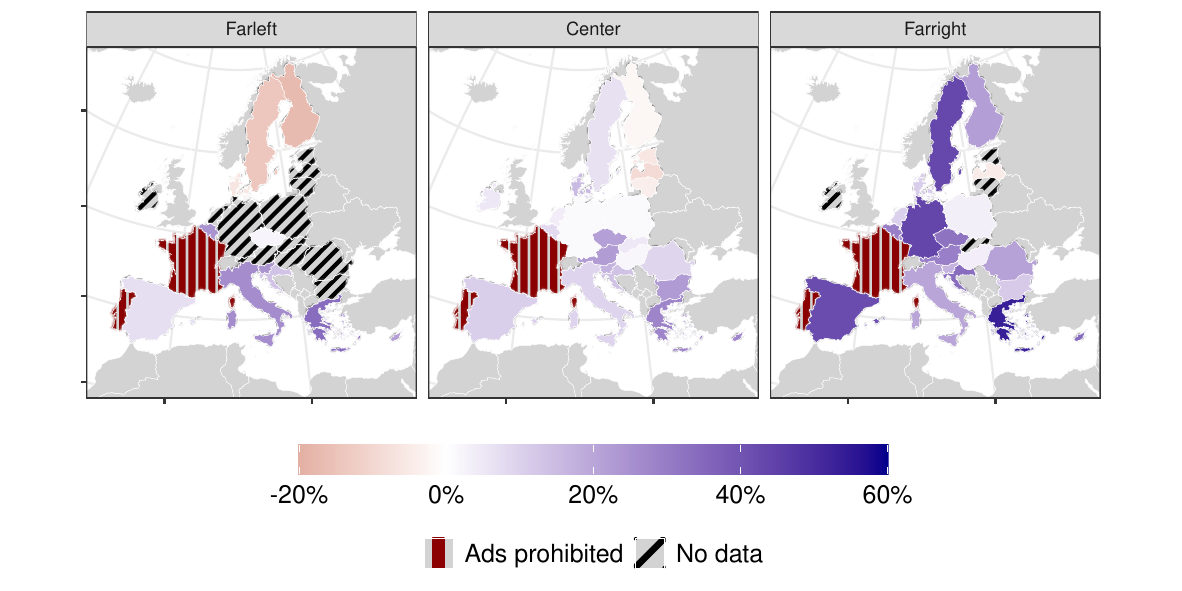}
\end{minipage}
    \caption{\textbf{Excess share of unique views by male users for ads from far-left, centrist, and far-right parties across the EU.} \figletter{a)},~Share of unique views by female and male users across the members of the European Union for ads published across far-left, center, and far-right parties. \figletter{b)},~The map displays the excess share of unique views by male users for political ads published on Facebook and Instagram by far-left, centrist, and far-right parties in European Union member states. A value near zero indicates a gender-balanced audience, while positive values reflect a higher proportion of male viewers. Darker shades represent countries where ads from extremist parties received significantly more unique views by male users compared to those from non-extremist parties. Party classifications are based on the PopuList database~\cite{Rooduijn.2024} (center is for reference). France and Portugal (hatched in red) prohibit political advertising during election campaigns. Countries hatched in black had no far-left or far-right party represented in the previous European Parliament or elected in the 2024 elections.}
    \label{fig:map_excess_male_extremist}
\end{figure}

\Cref{fig:map_excess_male_extremist} presents the excess share of unique views by male users for ads from far-left, centrist, and far-right parties across the European Union. Ads from far-right parties tend to exhibit a stronger gender imbalance, skewed toward male audiences. In contrast, ads from far-left parties show a tendency to reach more female than male users. Notable examples are Sweden and Finland, where ads from far-left parties are more likely to be viewed by female users (Sweden: 56\%; Finland: 57,\%), while far-right parties in these countries tend to reach predominantly male audiences (Sweden: 71\%; Finland: 61\%).

\subsubsection*{Regression analysis}

To examine the relationship between gender-based discrepancies in the audience of political ads and the political ideology of the advertising party (\ie, far-left, centrist, or far-right), we again rely on regression analysis. Specifically, we re-estimate the model from our main analysis, replacing the populist party indicator with two binary variables: one indicating whether an ad was published by a far-right party ($1$ if far-right, $0$ otherwise), and one indicating whether it was published by a far-left party ($1$ if far-left, $0$ otherwise). The reference category in this specification is ads published by centrist parties.

Following the approach from our main analysis, we again control for (1)~ad content, (2)~ad characteristics, and (3)~targeting strategies, and we include country fixed effects to account for country-specific heterogeneity. This specification allows us to isolate the association between algorithmic ad delivery and gender-based discrepancies across ads from the whole political spectrum.

\begin{figure}[H]
    \centering
    \includegraphics[width=0.8\linewidth]{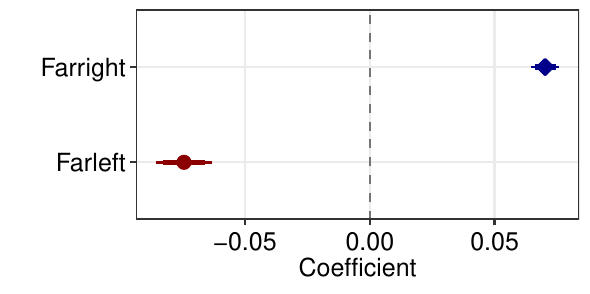}
    \caption{\textbf{The role of political extremism in gender-based discrepancies of political ads on social media.} Estimated coefficients from our regression model examining gender-based discrepancies between political ads published by far-right, far-left, or centrist parties. The coefficients capture the association between party affiliation and the excess share of unique views by male users. We find a positive and statistically significant coefficient for ads published by far-right parties, and a negative and statistically significant coefficient for far-left parties. This suggests that, all else equal, ads from far-right parties tend to attract significantly more male viewers, whereas far-left parties tend to have a larger female audience compared to centrist parties on average. Reported are the estimates (dot, diamond), along with $\alpha = 95\,\%$ confidence intervals (thick bars) and $99\,\%$ confidence intervals (thin bars). 
    }
    \label{fig:results_regression_extremist}
\end{figure}

\Cref{fig:results_regression_extremist} presents the estimated coefficients of the regression model. We find a negative and statistically significant coefficient for ads published by far-left parties (coefficient: $-0.074$, $\text{SE} = 0.004$, $t = -17.225$, $P<0.001$, 95\,\%~CI~$[-0.086, -0.063]$). This result indicates that, all else equal, ads published by far-left parties are associated with a 7.4 percentage point lower excess male audience share relative to centrist parties. In contrast, we find a positive and statistically significant coefficient for ads published by far-right parties (coefficient: $0.070$, $\text{SE} = 0.002$, $t = 32.665$, $P<0.001$, 95\,\%~CI~$[0.065, 0.075]$), indicating that these ads receive a 7.0 percentage point higher excess male audience share relative to centrist parties. Together, these findings highlight that ads from parties at both ideological extremes are associated with gender-skewed audiences on Facebook and Instagram, even after accounting for ad content, ad characteristics, targeting strategies, and country-specific heterogeneity.

\clearpage
\section*{Discussion}
\label{sec:discussion}


This study provides large-scale, cross-national evidence that the algorithmic delivery of political advertising on social media was systematically associated with gender-based discrepancies during the 2024 European Parliament elections. Across 25 EU member states, ads from populist parties---and particularly those at the far right of the ideological spectrum---are disproportionately delivered to male audiences, even when advertisers do not explicitly target by gender and after accounting for ad content, platform-level dynamics, and targeting strategies; all else equal, such ads are associated with a 6.2 percentage point higher excess male audience share.





Our findings align with a growing body of research that shows how algorithmic ad delivery can amplify demographic skews, including gender-based disparities documented in job, housing, and political advertising contexts \cite{Lambrecht.2019,Speicher.2018,Ali.2021,Imana.2021,Sankaranarayanan.2023}. We extend this literature by demonstrating that advertiser ideology itself is systematically associated with differential delivery outcomes, beyond explicit targeting choices or ad content. Unlike earlier studies that rely on single-country analyses or controlled experimental designs, our study leverages real-world campaign data from a large-scale supranational election, thereby strengthening the external validity of existing evidence. By extending prior single-country case studies \cite{Lambrecht.2019,Speicher.2018,Ali.2021,Imana.2021,Bar.2024b} to a continental electoral context, our results suggest that gender-based disparities in political ad delivery are not isolated anomalies but reflect a structural feature of algorithmic optimization in contemporary political campaigns.



A plausible mechanism underlying the observed gender-based discrepancies is that ad delivery algorithms optimize toward historical engagement patterns, which may be more pronounced among male users for populist, and particularly far-right, political content. Rather than reflecting advertiser intent, such dynamics are likely driven by learned correlations between political ideology, rhetorical style, and user behavior that are embedded in platform optimization processes \cite{ernst2019populists,Dai.2022,Capozzi.2023}. As a consequence, even ads that are explicitly gender-neutral may become demographically skewed through feedback loops in which early engagement by specific user groups influences subsequent delivery decisions. Prior work has shown that Facebook’s ad delivery systems can disproportionately favor male audiences even in non-political contexts \cite{Lambrecht.2019}, suggesting that similar mechanisms may operate in political advertising and, in turn, constrain women’s opportunities to engage equally with political information on social media.




Unequal exposure to political advertising on social media raises important normative concerns for democratic elections, as it undermines the principle of equal access to political discourse. When algorithmic delivery systematically skews political messaging toward specific demographic groups, voters may be differentially exposed to political viewpoints, limiting their ability to engage with the full spectrum of electoral choices. Gender-based disparities in ad delivery are particularly problematic given existing gaps in political participation and representation, as they may further marginalize groups that are already underrepresented in the political arena. In high-stakes elections, even modest differences in exposure can shape issue salience, mobilization, and ultimately vote choice \cite{Bar.2025}. Consistent with previous evidence on the role of social media in influencing electoral processes \cite{Aral.2019,LorenzSpreen.2022,Kim.2026}, these platforms may amplify the democratic consequences of algorithmic asymmetries in political communication.


Our results underscore important limitations in current transparency mechanisms for political advertising on social media. While platforms provide aggregate statistics on ad audiences, such information is insufficient to meaningfully assess how delivery decisions are made in practice or to audit the role of pricing, engagement signals, and auction dynamics in shaping exposure. The unexplained variation in ad delivery---even after accounting for observable content, contextual factors, and targeting---suggests that key components of platform behavior remain opaque. This lack of transparency constrains independent scrutiny by researchers and regulators and limits effective democratic oversight of algorithmic systems that play an increasingly central role in electoral communication. Notably, the stronger male skew observed for far-right parties and the corresponding female skew for far-left parties indicate that algorithmic delivery systems may internalize and amplify long-standing sociopolitical cleavages. These patterns mirror well-documented gender gaps in political participation and ideological alignment, raising concerns that platform optimization processes may reinforce, rather than mitigate, existing representational inequalities. In light of the growing electoral support for far-right parties across several EU member states and other democratic systems worldwide, such dynamics need careful attention. Moreover, given that populist and extremist parties often rely on emotionally-charged and polarizing messaging, algorithmic amplification of such content may further intensify political polarization, thereby underscoring the need for transparency and accountability in platform governance.


Our findings have direct implications for policymakers and social media platforms, particularly in the context of evolving regulation of political advertising in the EU and other parts of the world. Policymakers should consider strengthening disclosure requirements for political ads to include information on delivery mechanisms, engagement metrics, and auction dynamics, which are currently opaque \cite{Edelson.2020} yet central to algorithmic optimization. Regulatory approaches discussed at the EU level, such as randomized ad delivery within defined target populations, quota-based exposure rules, or separate auctions for political and commercial advertising, may help mitigate discriminatory delivery outcomes and reduce feedback-driven biases \cite{Turkel.2022,CounciloftheEuropeanUnion.2023}. In parallel, platforms themselves should implement systematic bias audits of political ad delivery, especially during election periods, to identify and correct demographic skews in real time. Together, mandatory transparency efforts and proactive platform governance could support more equitable exposure to political information while preserving the operational autonomy of advertising systems.


As with other research, ours is not free from limitations that present important directions for future work. First, our analysis focuses on political advertising on Facebook and Instagram, which together account for roughly 90\% of the social media market share in Europe \cite{StatcounterGlobalStats.2025}, making them central platforms for political communication during elections; nevertheless, future work should examine whether similar delivery patterns arise on other platforms such as TikTok, YouTube, or X, where user demographics and recommendation systems differ. Second, although we control for an extensive set of covariates related to ad content, platform-level characteristics, and targeting strategies, the observational nature of the data precludes causal claims about algorithmic bias or the mechanisms driving delivery outcomes. This issue mirrors a broader one of social media research in that our analysis is constrained by the limited transparency of platform-provided data: the Meta Ad Library offers only coarse information on audience composition and does not disclose pricing dynamics, engagement signals, or the internal logic of delivery algorithms \cite{Edelson.2020}, thereby limiting our ability to disentangle optimization incentives from algorithmic behavior. Finally, while our study centers on ad delivery, we do not examine broader communication strategies, such as message framing, narrative choices, or campaign coordination across ads; future interdisciplinary work combining computational methods and political communication theory could provide a more comprehensive understanding of how parties deploy social media advertising beyond targeting alone.


Overall, our findings demonstrate that algorithmic ad delivery plays a central role in shaping political exposure on social media, beyond advertisers’ explicit intent. In the absence of stronger transparency and accountability, such systems risk systematically disadvantaging certain demographic groups and narrowing access to political discourse. Ensuring fairness in political advertising, therefore, requires more efforts to safeguard equal access to political information in democratic elections.




\clearpage
\section*{Materials and Methods}
\label{sec:methods}

\subsection*{Data}

\subsubsection*{Social media ads for political advertising}

Our analysis relies on a large-scale dataset of over \num{110000} political ads published by 968 candidates and 453 parties on Meta Platforms, namely Facebook (\url{http://facebook.com}) and Instagram (\url{http://instagram.com}), across 25 member states of the European Union during the 2024 European Parliament election. We focus on Meta due to its dominant market share of approximately 90\,\% in Europe \cite{StatcounterGlobalStats.2025} and its central role for political communication \cite{Bond.2012, Lalancette.2019, Bar.2025, Bar.2024b}. 
In particular, Meta enables highly targeted political advertising at a relatively low cost and is thus considered an essential tool for modern election campaigns \cite{Bar.2024b, Bar.2025, Fowler.2021}. 

We collected ads published during the three months leading up to the election, that is, from March 9 to election day on June 9, 2024. The European Election takes place over several days, from June 6 to June 9, 2024, to accommodate national voting traditions. For example, the Netherlands votes on June 6, while most countries vote on June 9 \cite{Bundeswahlleiter.2024}. This observation period was chosen to reflect the main phase of the election campaign. In Germany, for instance, candidacies were officially announced on March 18. 
To test the robustness of our findings, we also repeated the analysis using alternative observation periods (see \Cref{supp:robustness_periods}).

We collected political advertisements using the Meta Ad Library API (\url{http://facebook.com/ads/library}), a public archive that includes all political ads published on Facebook and Instagram \cite{Meta.2022}. For each ad, the Ad Library provides detailed information, including: (1)~ad content, (2)~page name, (3)~ad spending (in EUR), (4)~sponsor, (5)~number of impressions, and (6)~number of unique views. Impressions refer to the total number of times an ad is displayed on any user's screen, while unique views indicate how many users saw the ad at least once. The latter serves as the basis for our main dependent variable in the regression analysis. The Meta Ad Library also specifies whether an ad appeared (1)~only on Facebook (referred to as \emph{Facebook-only}), (2)~only on Instagram (\emph{Instagram-only}), or (3)~on both platforms simultaneously (\emph{Dual platform}). In addition, the Ad Library reports the distribution of impressions and unique views by age group (\ie, 18--24, 25--34, 35--44, 45-–54, 55-–64, 65+) and gender (\ie, female, male, diverse).

Media content plays a central role in digital advertising on social media platforms \cite{Bar.2023b, Totti.2014}. To analyze this aspect, we further collected all images and videos associated with political ads from the Meta Ad Library using the AdDownloader Python package \cite{Gitu.2024}. In total, we retrieved 75916 images and 94688 videos corresponding to ads included in our sample.

We further used the \emph{Meta Ad Targeting Dataset}~\cite{Meta.2023b} to access targeting information for each ad. The Meta Ad Targeting Dataset contains targeting data for all social issues, electoral, and political ads published on Facebook or Instagram since August 3, 2020~\cite{Meta.2023b}. Specifically, we queried information on the demographics (age and gender) of users targeted by each ad. Additional targeting categories such as interests, behaviors, job titles, and locations are not available via the Meta Content Library. 


\subsubsection*{Party and candidate information}
\label{met:matching}
To identify the candidate or party publishing an ad, we rely on two resources: (1)~During the 2024 European elections, Meta published a list of verified accounts associated with political candidates in each EU member state. The online activity of these accounts was tracked through a public dashboard provided by Crowdtangle, owned by Meta Platforms and discontinued in August 2024 \cite{Buijs.2024}. (2)~We compiled a list of all candidates running in the 2024 European elections from official national election records. From this list, we also inferred all participating parties and identified their associated European political groups. Together, these sources provide a comprehensive set of candidates and parties that may have used Facebook and Instagram for their election campaigns.

We then matched each ad to these resources based on the ad's page name and sponsor. This approach allows us to identify ads that (1)~were sponsored by a specific candidate or party but published under a different page name (\eg, candidates sponsoring political ads for the regional chapter of their party, or vice versa), and (2)~were not sponsored by the candidate or party, yet published on a candidate's or party's page on Facebook or Instagram (\eg, a party sponsoring a political ad for their candidate, or an electoral coalition comprising multiple parties\footnote{Parties in Italy and Spain, for example, often form electoral coalitions and run with a joint list for the election}). 
This matching procedure ensures a comprehensive sample of all political ads associated with a candidate or party.

Our matching process followed a hierarchical three-step procedure: (i)~we first searched for exact matches between the ad's page name or sponsor and the candidate or party name. (ii)~If no exact match was found, we checked whether the candidate or party name was contained within the page name or sponsor (\eg, a page name like ``John Doe -- European Candidate'' is matched to the candidate name ``John Doe''); (iii)~for all remaining cases, we applied fuzzy string matching using the \texttt{WRatio} function from the \textit{rapidfuzz} Python library. This function computes a similarity score between two strings by combining Levenshtein distance with additional measures such as partial ratio, token sort ratio, and token set ratio, providing a robust and comprehensive assessment of their similarity. We used a conservative similarity threshold of 0.95 to minimize false positives, as experiments with lower thresholds resulted in considerable mismatches.

The procedure strictly prioritizes exact matches over substring matches, and substring matches over similarity-based matches. In cases of multiple potential matches, we manually verified which candidate or party name was correct. Additionally, we manually reviewed unmatched accounts from the candidate list published by Meta to ensure completeness.


\subsubsection*{Identifying populist parties}

The main explanatory variable in our analysis is whether a party is classified as ``populist'' or not. To identify ads by populist parties in our sample, we rely on the ``PopuList,'' a widely used resource that categorizes European parties based on their political stance~\cite{Rooduijn.2024}. The PopuList was developed through a collaboration between researchers from the University of Amsterdam, the University of York, the Fondation Nationale des Sciences Politiques, and the Nederlandse Organisatie voor Wetenschappelijk Onderzoek, in partnership with The Guardian and the ECPR Standing Group on Extremism and Democracy. It has been frequently used in prior work to determine the political affiliation of European parties~\cite{Capozzi.2023, Zulianello.2020, Norris.2020}. In addition to populist classification, we also retrieve each party’s ideological orientation as ``far-right,'' ``center,'' or ``far-left,'' which we use later as part of our robustness checks (see \Cref{supp:robustness_checks}). For ads published by individual candidates, we infer their political affiliation based on the party with which they are officially associated.

Political competition in Europe is characterized by a diverse multi-party system. 
In many countries, however, elections are dominated by a few major parties. To reflect this structure, we focus our analysis on the major parties participating in the 2024 European Elections. We define a party as ``major'' if it was either represented in the previous European Parliament or successfully elected in the 2024 European elections. Examples of major parties include the German ``SPD,'' the Spanish ``Partido Popular,'' and the Italian ``Partito Democratico.'' 


\subsection*{Statistical analysis}

We use regression analysis to compare the gender distribution of audiences for political ads from populist and non-populist parties published on Facebook and Instagram during the 2024 European Parliament elections. Specifically, we estimate a linear regression model in which the main dependent variable is the excess share of unique views by male users of an ad (\emph{Excess male share}). Specifically, \emph{Excess male share} is the difference between the share of unique views by male vs. female users. This information is provided directly by Meta Ad Library. 

Let $y_i$ denote the excess share of unique views by male users for ad $i$, $\phi_i$ be a binary variable indicating whether the ad was published by a populist party ($\phi_i = 1$ if populist, $0$ otherwise), and $X_i$ be a vector of control variables. The regression model is specified as follows:
\begin{equation}
    y_i = \alpha + \beta \phi_i + \gamma X_i + \delta_i + \varepsilon_i,
    \label{eqn:regression_model}
\end{equation} 
where $\alpha$ represents the intercept, $\beta$ captures the association between publication by a populist party and the excess share of male views in percentage points, $\gamma$ denotes the coefficients associated with the control variables in $X_i$, $\delta_i$ are country-fixed effects, and $\varepsilon_i$ is the residual.

We estimate the model using Ordinary Least Squares (OLS) and test whether the coefficients are significantly different from zero using two-sided $t$-tests. Standard errors are clustered at the national party level to account for potential dependencies among ads published by the same party. The coefficient of interest, $\beta$, is interpreted as follows: if an ad is published by a populist party (\ie, $\phi_i = 1$), all else equal, the share of unique views by male users is predicted to increase by $\beta\times100\%$ percentage points on average.

For our regression model, we include a comprehensive set of control variables (\ie, $X_i$ in Eq.~\ref{eqn:regression_model}) along with country-level fixed effects (\ie, $\delta_i$ in Eq.~\ref{eqn:regression_model}) to account for a wide range of factors that may influence ad delivery on social media platforms. Specifically, we control for (1)~the content of the ad, (2)~the targeting strategy employed by the advertiser, (3)~ad-level characteristics such as timing and platform-level competition, and (4)~country-level heterogeneity through fixed effects. These controls help isolate the association between ad sponsorship by populist parties and gender differences in ad delivery.
 
\emph{(1)~Ad content:} Ad content plays an important role in the delivery of ads on social media platforms. Prior research has shown that certain types of content are disproportionately delivered to users based on demographic characteristics \cite{Imana.2021, Imana.2024, Ali.2019, Ali.2021, Ali.2023}. For instance, STEM career ads have been found to be more frequently shown to male users than to female users~\cite{Lambrecht.2019}. To account for audience skew related to ad content, we control for both the textual and visual components of each ad in our sample.

To capture the textual content of an ad, we analyze its sentiment and classify each ad as conveying either ``positive,'' ``neutral,'' or ``negative'' sentiment. Specifically, we first translate non-English ad texts into English using Google Translate. We then apply a RoBERTa-based transformer model fine-tuned for sentiment analysis on social media text (Hugging Face model card: \url{cardiffnlp/twitter-roberta-base-sentiment-latest})~\cite{Barbieri.2020}. The model was trained on approximately 124 million social media posts and is well-suited for detecting nuanced sentiment in short-form text such as social media ads.

To process the media content (\ie, images and videos) of each ad, we use DINOv2 (Hugging Face model card: \url{facebook/dinov2-base}), a state-of-the-art vision transformer developed by Meta and pretrained on over 142~million images~\cite{Oquab.2023}. For ads containing static images, we extract image embeddings by applying DINOv2 and using the pooled output from its final layer, yielding a fine-grained visual representation of each ad image. For video content, we extract the thumbnail (\ie, the first frame) along with up to 15 equally spaced frames, sampled at a rate of 2 frames per second. This strategy captures both the initial visual impression of the ad—as shown when scrolling—and the broader visual content across the full video. We then compute embeddings for each extracted frame using DINOv2 and average the pooled outputs to obtain a single video representation per ad. To address issues of multicollinearity associated with including high-dimensional content embeddings in our regression models, we apply Principal Component Analysis (PCA). Specifically, we reduce the image embeddings to 213 components, which retain 80\% of the total variance in the original embeddings and preserve most of the visual information. As a robustness check, we repeated the analysis using 371 components and found consistent results.

\emph{(2)~Ad characteristics:} Ad characteristics are likely to influence the composition of the audience of political ads. To account for this, we include control variables for ($i$)~competition, ($ii$)~timing, ($iii$)~the platform on which the ad was published, and ($iv$)~the publisher of the ad. ($i$)~\emph{Competition}: Political ads published concurrently may compete for user attention, potentially shaping the observed audience composition \cite{Bossetta.2018}. To control for this, we include the number of concurrently active political ads in a given country, capturing the level of platform-level competition. ($ii$)~\emph{Timing}: Temporal features such as the day of the week and ad duration can affect the reach and user engagement with political ads~\cite{Salonen.2024, Bar.2024b}. We thus control for the weekday on which an ad was launched and the total number of days an ad remained active. ($iii$)~\emph{Platform}: Platform-specific dynamics, including differences in user demographics and behavior, may influence the reach and delivery of political ads~\cite{Bossetta.2018}. We control for this by including dummy variables indicating whether an ad was published on Facebook, Instagram, or both platforms simultaneously. ($iv$)~\emph{Publisher}: In the context of the European elections, both parties and individual candidates may pursue distinct advertising strategies that influence the gender distribution of ad audiences. For instance, while a party may generally attract a predominantly male audience, a specific candidate may resonate more strongly with female voters. To account for this, we include a binary variable indicating whether an ad was published by a candidate ($=1$ if yes, $=0$ otherwise).

\emph{(3)~Targeting strategy:} The targeting strategy of an ad determines the audiences that social media platforms are expected to deliver the ad to. Different targeting strategies are likely to influence the observed composition of the audience.
We control for age-based targeting by including the share of targeted users across different age groups (\ie, 18--24, 25--34, 35--44, 45--54, 55--64, and 65+). This follows the rationale that the gender distribution of Facebook and Instagram users varies across age groups. For example, in Spain, users aged 25--34 are almost equally likely to be male (10.9\,\%) or female (11.1\,\%), whereas older users aged 55--64 are more likely to be female (7.0\,\% female vs. 5.6\,\% male) \cite{NapoleonCat.2024}. Thus, if an advertiser targets older users in Spain, their audience is more likely to skew female.

\emph{(4)~Country-level heterogeneity:} Ad delivery may vary across countries within the European Union. For example, national regulations may require platform-specific adjustments to algorithmic ad delivery. Likewise, the gender distribution of Facebook and Instagram users may differ across countries, which could influence observed audience patterns. To account for such country-level heterogeneity, we include country-fixed effects in our regression model. Specifically, we introduce a set of dummy variables indicating the country in which each ad was published ($=1$ if the ad was published in a given country, $=0$ otherwise). These fixed effects control for any unobserved, country-specific factors that are constant across ads within the same country.

Sentiment, weekday, and platform are multi-level categorical variables. To include them in our regression model, we define reference categories and interpret coefficients relative to these baselines. Specifically, we use ``neutral'' sentiment, ``Monday'' as the weekday, and ``both platforms'' as the platform reference category. We assessed multicollinearity using variance inflation factors (VIFs). All ad covariates show low multicollinearity, with VIF values below 1.4.

\subsection*{Robustness checks}
\label{sec:robustness_checks}

To evaluate the robustness of our findings, we conduct a series of additional analyses. Specifically, we evaluate how our findings are robust to (1)~alternative definitions of our dependent variable, (2)~model specification, (3)~data sample, (4)~modeling approaches, and (5) different covariate selections. Detailed results are presented in \Cref{supp:robustness_checks}.

(1)~Alternative dependent variable: We assess whether our findings are robust to alternative definitions of the dependent variable. In the main model, we use the excess share of unique views by male users as the dependent variable. As a robustness check, we re-estimate our model simply using the share of unique users by male users.

(2)~Model specification: We test the sensitivity of our results to changes in model specification. In particular, we add the country-level share of male users on Facebook and Instagram as an additional control to account for differences in baseline gender distribution across national user populations. We collect the actual gender distribution of users on Facebook and Instagram in June 2024 for each country from NapoleonCat (\url{https://napoleoncat.com}). This check addresses the concern that observed gender discrepancies in ad delivery may reflect underlying platform demographics rather than algorithmic bias. However, including this variable introduces perfect multicollinearity with the country-fixed effects, and we therefore omit them in this specification. The motivation to include them in the main analysis is that they capture not only baseline gender shares but also other country-specific heterogeneity (e.g., regulations).

(3)~Data sample: We explore how different observation periods affect our findings. To do so, we restrict the sample to ads published during (a)~the final two months and (b)~the final month of the campaign period. This accounts for potential shifts in audience composition or platform dynamics closer to the election.


(4)~Modeling approach: We implement a non-parametric robustness check using propensity score matching (PSM). Specifically, we match ads from populist and non-populist parties based on observed covariates, including ad content, targeting strategies, and ad-level characteristics.
Additionally, we re-estimated our baseline models using a mixed-effects framework. Specifically, we fitted a linear mixed-effects model where we use random intercepts for each country, rather than fixed effects, while retaining our full set of ad content, targeting, and platform controls

(5)~Covariates ablation: We explore how the exclusion of different covariates presented in the main analyses changes the effect. Specifically, we remove the controls for age targeting and media content.
We notice that, while the coefficient changes in magnitude, the presented effects are still significant.

Overall, we found a robust and statistically significant coefficient for ads by populist parties. In particular, our results are robust across different model specifications, data samples, and estimation techniques.

\subsection*{Additional analyses: The role of political extremism}

Populist parties may originate from both ends of the political spectrum~\cite{Vachudova.2021}. For example, in Southern Europe, left-leaning populist parties have gained prominence, whereas in Central and Eastern Europe, populist movements tend to be aligned with the political right~\cite{Vachudova.2021}. As a result, algorithmic delivery of political ads may not only differ between populist and non-populist parties, but also between the ideological fringes of the spectrum, \ie, the far-right and the far-left.

We thus conduct an additional analysis in which we modify our main explanatory variable to examine whether gender discrepancies in ad delivery vary across the ideological spectrum. Specifically, we re-estimate our main regression model, replacing the populist party indicator with two binary variables: one indicating whether an ad was published by a far-right party ($=1$ if far-right, $=0$ otherwise), and one indicating whether it was published by a far-left party ($=1$ if far-left, $=0$ otherwise). The reference category in this specification is ads by centrist parties. To identify the ideological orientation of each party, we again rely on the classification in the PopuList \cite{Rooduijn.2024}.


To further disentangle the intersecting roles of populist rhetoric and political extremism in algorithmic ad delivery, we conducted an interaction analysis.
While our primary models evaluate populism and political ideology as independent predictors, populist parties in Europe span both the far-left and far-right of the political spectrum. 
To account for this, we extended our baseline linear regression model by introducing an interaction term between the populist indicator and the political extremism variable. 
This specification allows us to assess whether the male-skewing effect of populist messaging operates uniformly or is amplified or dampened depending on a party's underlying ideological orientation. 
Consistent with our main approach, this model controls for ad content, targeting strategies, ad characteristics, and country-fixed effects, with standard errors clustered at the national party level.
The detailed estimation results for this interaction model are reported in \Cref{supp:interaction}.


\clearpage
\section*{Additional information}

\vspace{0.4cm}
\noindent
\textbf{Code availability.} All code to replicate our analyses will be made available upon publication in a public repository.

\vspace{0.4cm}
\noindent
\textbf{Data availability.} All data used for the analysis is available across different sources. Data on political ads on Facebook and Instagram is available via the Meta Ad Library: \url{https://www.facebook.com/ads/library}. Targeting data is available via the Meta Ad Targeting Dataset: 
\url{https://developers.facebook.com/docs/fort-ads-targeting-dataset}. To ensure reproducibility, we provide ids for all ads in our dataset, which can be used to retrieve the original data through the Meta Ad Library (both through the API and with the web interface by simply using the id as query). Election data is available via the official national election records. The raw data can be retrieved from D.B. upon reasonable request to protect the privacy of the candidates.

\section*{Author information} 

\textbf{Author contributions.} D.B., F.P., G.D.F.M., and S.F. contributed to conceptualization. D.B., F.C., and F.P. contributed to data analysis. D.B., F.C., F.P., G.D.F.M., and S.F. contributed to results interpretation and manuscript writing. D.B., F.C., F.P., G.D.F.M., and S.F. approved the manuscript.


\vspace{0.4cm}
\noindent
\textbf{Competing interests.} The authors declare no competing interests.



\clearpage
\bibliography{literature}


\clearpage 
\appendix
\label{sec:supplementary}
\renewcommand{\thetable}{S\arabic{table}}
\renewcommand{\thefigure}{S\arabic{figure}}
\setcounter{figure}{0}
\setcounter{table}{0}    

\begin{minipage}[t]{\textwidth}
\begin{center}
\huge\bfseries Supplementary Materials
\end{center}
\vspace{1cm}
\end{minipage}

\tableofcontents

\clearpage

\section{The 2024 European Elections}
\label{supp:politics_EU}

The 2024 elections to the European Parliament constitute one of the largest democratic exercises globally. The European Parliament is the only directly elected institution of the European Union (EU) and serves as its legislative body, sharing law-making, budgetary, and supervisory powers with the Council of the European Union. In the 2024 elections, a total of 720 Members of the European Parliament (MEPs) were elected to represent more than 450 million citizens across the EU’s 27 member states (\url{https://results.elections.europa.eu/en}).

European Parliament elections are held every five years and take place simultaneously across all member states, although voting occurs on different days depending on national traditions (\url{https://elections.europa.eu/en/how-elections-work}). In 2024, elections were held between June 6 and June 9, with some countries voting earlier (e.g., the Netherlands on June 6) and most others voting on June 9. Despite the supranational nature of the election, electoral rules are largely determined at the national level. Each member state is allocated a fixed number of seats that is broadly proportional to its population, subject to the principle of degressive proportionality, whereby smaller states receive a higher number of seats per capita than larger ones.

Elections to the European Parliament are conducted using proportional representation systems (\url{https://elections.europa.eu/en/how-elections-work}), though the specific design varies across countries. Depending on national law, voters may cast a vote for a closed party list, an open list allowing preference votes for individual candidates, or—in some cases—single transferable votes. Unlike many national parliamentary elections, European Parliament elections do not generally involve single-member constituencies; instead, multi-member districts or nationwide constituencies are used (\url{https://www.europarl.europa.eu/factsheets/en/sheet/21/elections-to-the-european-parliament}).

Political competition in European Parliament elections is characterized by a multi-level structure. Candidates typically run as members of national political parties, which in turn are affiliated with transnational political groups within the Parliament, such as the \textit{European People’s Party} (EPP), the \textit{Progressive Alliance of Socialists and Democrats} (S\&D), \textit{Renew Europe} or the \textit{ Greens/European Free Alliance}. As a result, campaigns are shaped by both domestic political considerations and broader European-level issues, including economic governance, migration, climate policy, foreign policy, and democratic norms.

Recent European Parliament elections have also been characterized by the growing electoral strength of populist and far-right parties in several member states (see the European Parliament Think Tank; ``Stock-taking of the 2024 European Parliament Elections''; URL: \url{https://www.europarl.europa.eu/thinktank/en/document/IUST_BRI(2025)771520}). This trend has been documented both within the EU and in comparative research on democratic systems globally, thereby raising concerns about political polarization, democratic backsliding, and the role of digital media in electoral competition (see the European Parliament Think Tank; ``Social media platforms and challenges for democracy, rule of law and fundamental rights''; URL: \url{https://www.europarl.europa.eu/thinktank/en/document/IPOL_STU%282023%29743400}; and ``Youth and social media''; URL: \url{https://www.europarl.europa.eu/thinktank/en/document/EPRS_BRI%282025%29779235}). Against this backdrop, political campaigning has increasingly shifted toward online platforms, with social media playing a central role in voter mobilization, issue framing, and political advertising.

An exception in France, which imposes strict legal restrictions on paid political advertising in the period immediately preceding elections. French law prohibits political advertising on broadcast media and imposes strong limitations on online political advertising during the official campaign period (\url{https://www.legifrance.gouv.fr/jorf/id/JORFTEXT000047062308}), which substantially reduces the volume of paid political ads observable on platforms such as Facebook and Instagram. As a consequence, political advertising activity from France is only partially captured in our dataset and does not fully reflect campaign dynamics during the final and most salient phase of the election (VoxEurope and Civil Liberties Union for Europe: ``Elections Monitoring 2024: France. Electoral Integrity and Political Microtargeting in the European Parliament Elections: An Evidence-Based Analysis''; URL: \url{https://www.liberties.eu/f/hstorx}). 

The scale, diversity, and simultaneity of the 2024 European Parliament elections make them a particularly informative setting for studying political communication on social media. The election combines heterogeneous national electorates, multiple party systems, and a shared digital advertising infrastructure operated by global platforms, offering a unique opportunity to examine how algorithmic ad delivery operates across countries within a single, high-stakes democratic contest.

\newpage 

\section{Robustness checks}
\label{supp:robustness_checks}

\subsection{Alternative dependent variable}

We use an alternative dependent variable: the share of unique users by male users for the populist vs non-populist ads and the two classes of political extremism.
These analysis resulted again in a negative and statistically significant coefficient for ads published by far-left parties (coefficient: $-0.039$, $\text{SE} = 0.002$, $t = -18.007$, $P<0.001$, 95\,\%~CI~$[-0.043, -0.034]$), and a positive and statistically significant coefficient for ads published by far-right parties (coefficient: $0.035$, $\text{SE} = 0.001$, $t = 33.008$, $P<0.001$, 95\,\%~CI~$[0.034, 0.037]$). 
The analysis also resulted in a positive and significant coefficient for ads published by populist parties  (coefficient: $0.062$, $\text{SE} = 0.017$, $t = 3.51$, $P<0.001$, 95\,\%~CI~$[0.027, 0.097]$).

\subsection{Alternative model specification}

We use an alternative specification of the regression model. 
We added the country-level share of male users on Facebook and Instagram, and, in order to avoid multicollinearity, we exclude the country fixed effects.
These analysis resulted again in a negative and statistically significant coefficient for ads published by far-left parties (coefficient: $-0.034$, $\text{SE} = 0.004$, $t = -8.007$, $P<0.001$, 95\,\%~CI~$[-0.042, -0.026]$), and a positive and statistically significant coefficient for ads published by far-right parties (coefficient: $0.0585$, $\text{SE} = 0.002$, $t = 28.808$, $P<0.001$, 95\,\%~CI~$[0.054, 0.062]$). 
The analysis also resulted in a positive and significant coefficient for ads published by populist parties  (coefficient: $0.053$, $\text{SE} = 0.002$, $t = 27.93$, $P<0.001$, 95\,\%~CI~$[0.0495, 0.0570]$).

\subsection{Data sample restrictions}
\label{supp:robustness_periods}

To ensure our findings are not driven by early-campaign dynamics and to account for potential shifts in audience composition closer to the election day, we tested the sensitivity of our results to different observation periods. Specifically, we re-estimated our main analysis model after restricting the sample to ads published during (i) the final two months and (ii) the final one month leading up to the election. The association between populist parties and a male-skewed audience remained highly robust across both restricted timeframes: For (i), when limiting the data to the final two months of the campaign, the estimated effect of populist messaging on the male audience share was 6.3 percentage points (coefficient = $0.063$, SE = $0.002$, $t = 26.93$, $p < 0.001$, 95\% CI $[0.059, 0.068]$). For (ii), even when analyzing only the highly competitive final month of the campaign, the effect remained substantial and statistically significant at 5.8 percentage points (coefficient = $0.058$, SE = $0.003$, $t = 21.42$, 4, 95\% CI $[0.053, 0.063]$). These results closely mirror our baseline estimate of 6.2 percentage points, confirming that the observed gender-based discrepancies persist consistently throughout the peak phases of the electoral campaign.

\subsection{Modeling approaches}

To further ensure our findings are robust against potential selection bias from observable ad characteristics, we performed a propensity score matching (PSM)\cite{austin2011introduction} analysis. The procedure successfully matched all 34,080 populist ads (the treated group) against the pool of 76,271 non-populist ads (the control group), which corresponds to an effective sample size of 7,511 in the control group after weighting. The estimated average treatment effect (ATE) implies that populist messaging is associated with a statistically significant increase in male audience share. Specifically, compared to observationally equivalent non-populist ads, populist ads tend to increase the male audience proportion by approximately 4.56 percentage points (coefficient = $0.0456$, SE =$ 0.00167$, $z = 27.3$, $p < 0.001$, 95\% CI $[0.0423, 0.0489]$). The PSM estimate closely aligns with our main specification, thereby reinforcing the robustness of the conclusion that populist campaigns systematically skew toward male audiences.

In our main analysis, we account for country-level heterogeneity using country fixed-effects and handle dependencies among ads by clustering standard errors at the national party level. 
To ensure our findings are not an artifact of this specific modeling choice, we re-estimated our baseline models using a mixed-effects framework. Specifically, we fitted a linear mixed-effects model assigning random intercepts for each country, rather than fixed effects, while retaining our full set of controls. The estimated effect of populist messaging remained robust, statistically significant, and consistent in magnitude with our main model specification.
Under the mixed-effects specification, ads published by populist parties are associated with an approximate 6 percentage point increase in the male audience share (coefficient =$ 0.06$, SE = $0.002$, $t = 30.575$, $p < 0.001$). This confirms that the observed gender-based discrepancies hold regardless of how national-level variance is explicitly modeled.

\subsection{Sensitivity to covariate selection}

To evaluate the sensitivity of our findings to the inclusion of specific blocks of covariates, we performed a study where we systematically added/removed age targeting and visual media controls.

Our baseline specification accounts for platform, timing, sentiment, ad competition, and country fixed effects, but excludes age and media controls; here, populist messaging is associated with a 7.8 percentage point increase in the male audience share (coefficient = $0.078$, SE =$ 0.021$,$ p < 0.001$).
Introducing visual media controls to this baseline slightly attenuates the estimated effect to 5.5 percentage points (coefficient = $0.055$, SE = $0.019$, $p = 0.003$). 
Conversely, controlling for age targeting without media covariates yields an estimated effect of 8.5 percentage points (coefficient = $0.085$, SE =$ 0.020$,$  p < 0.001$).

For comparison, our full model, which incorporates both age targeting and media covariates concurrently, suggests an effect of at 6.2 percentage points (coefficient =$ 0.062$, SE =$ 0.0020$,$ p < 0.001$). Across all alternative specifications, the coefficient for populism remains positive and statistically significant. This confirms that the observed algorithmic gender skew is a fundamental characteristic of populist ad delivery, rather than a mere artifact of specific age-targeting strategies or visual media choices.

\section{Interaction between populism and political extremism}
\label{supp:interaction}

To explore whether the audience distortions driven by populism vary depending on a party's ideological fringe, we estimated a model interacting the populist indicator with the extremist variable. 
The results reveal nuanced dynamics.
We find that both populism and far-right ideology independently contribute to a male-skewed ad audience. 
The interaction term for populist far-right parties is negative and statistically significant (coefficient = $-0.189$, SE =$ 0.063$, $p = 0.003$). 
This sub-additive effect indicates that, while ads from populist far-right parties do strongly skew male, the combined algorithmic distortion is not simply the sum of its parts.
This points to a potential nuance in the platform's delivery algorithms, where combining two male-skewing characteristics does not proportionally compound the audience distortion. 
Conversely, the interaction between populism and far-left ideology is not statistically significant (coefficient = $-0.011$, SE = $0.069$, $p = 0.873$), indicating that populist traits do not significantly alter the delivery patterns of far-left ads beyond their baseline tendencies.
\end{document}